\documentclass{llncs}
\pdfoutput=1
\usepackage{llncsdoc}
\usepackage{amssymb}
\usepackage{graphicx}
\usepackage{subfigure}
\usepackage{amsmath}
\usepackage{graphics}
\usepackage{tabularx}
\usepackage{url}
\urldef{\mailsa}\path|{royaen, jeffk, alexandg, crandall}@cs.unm.edu|

\newenvironment{packed_item}{
\begin{itemize}
  \setlength{\itemsep}{1pt}
  \setlength{\parskip}{0pt}
  \setlength{\parsep}{0pt}
}{\end{itemize}}
\begin{document}
\begin{sloppypar}

\mainmatter  

\title{Detecting Intentional Packet Drops on the Internet via TCP/IP Side Channels:\\ {\it Extended Version}}
\titlerunning{Detecting Intentional Packet Drops on the Internet}

\author{Roya Ensafi \and Jeffrey Knockel \and Geoffrey Alexander \and Jedidiah R. Crandall}

\institute{Department of Computer Science, University of New Mexico, USA.\\\mailsa}


\maketitle
\let\thefootnote\relax\footnote{This is the extended version of a paper from the 2014 Passive and Active Measurements Conference (PAM), March $10^{th}$--$11^{th}$, 2014, Los Angeles, California.}
\begin{abstract}
We describe a method for remotely detecting intentional packet drops on the
Internet \emph{via} side channel inferences.  That is, given two arbitrary IP
addresses on the Internet that meet some simple requirements, our proposed
technique can discover packet drops (\emph{e.g.}, due to censorship) between
the two remote machines, as well as infer in which direction the packet drops
are occurring.  The only major requirements for our approach are a client with
a global IP Identifier (IPID) and a target server with an open port.  We
require no special access to the client or server.  Our method is robust to
noise because we apply intervention analysis based on an
autoregressive-moving-average (ARMA) model.  In a measurement study using our
method featuring clients from multiple continents, we observed that, of all
measured client connections to Tor directory servers that were censored, 98\%
of those were from China, and only 0.63\% of measured client connections from
China to Tor directory servers were not censored.  This is congruent with
current understandings about global Internet censorship, leading us to conclude
that our method is effective.

\end{abstract}

            
            


\section{Introduction} \label{sec:intro}
Tools for discovering intentional packet drops are important for a variety of
applications, such as discovering the blocking of Tor by ISPs or nation
states~\cite{torblog}.  However, existing tools have a severe limitation: they
can only measure when packets are dropped in between the measurement machine
and an arbitrary remote host.  The research question we address in this paper
is: can we detect packet drops between two hosts without controlling either of
them and without sharing the path between them?  Effectively, by using idle
scans our method can turn approximately 1\% of the total IP address space into
conscripted measurement machines that can be used as vantage points to measure
IP-address-based censorship, without actually gaining access to those machines.
We can achieve this because of information flow in their network stacks.

Antirez~\cite{antirez} proposed the first type of idle scan, which we call an
IPID idle port scan.  In this type of idle scan an ``attacker'' (which we will
refer to as the \emph{measurement machine} in our work) aims to determine if a
specific port is open or closed on a ``victim'' machine (which we will refer to
as the \emph{server}) without using the attacker's own return IP address.  The
attacker finds a ``zombie'' (\emph{client} in this paper) that has a global IP
identifier (IPID) and is completely idle.  In this paper, we say that a machine
has a global IPID when it sends TCP RST packets with a globally incrementing
IPID that is shared by all destination hosts.  This is in contrast to machines
that use randomized IPIDs or IPIDs that increment per-host.  The attacker
repeatedly sends TCP SYN packets to the victim using the return IP address of
the zombie, while simultaneously eliciting RST packets from the zombie by
sending the zombie SYN/ACKs with the attacker's own return IP address.  If the
victim port that SYN packets are being sent to is open, the attacker will
observe many skips in the IPIDs from the zombie.  Nmap~\cite{nmapbook} has
built-in support for antirez's idle scan, but often fails for Internet hosts
because of noise in the IPID that is due to the zombie sending packets to other
hosts.  Our method described in this paper is resistant to noise, and can
discover packet drops in either direction (and determine which direction).
Nmap cannot detect the case of packets being dropped from client to server
based on destination IP address, which our results demonstrate is a very
important case.

Two other types of idle scans were presented by Ensafi \emph{et
al.}~\cite{roya}, including one that exploits the state of the SYN backlog as a
side channel.  Our method is based on a new idle scan technique that can be
viewed as a hybrid of the IPID idle scan and Ensafi \emph{et al.}'s SYN backlog
idle scan.  Whereas Ensafi \emph{et al.}'s SYN backlog idle scan required
filling the SYN backlog and therefore causing denial-of-service, our technique
uses a low packet rate that does not fill the SYN backlog and is non-intrusive.
The basic insight that makes this possible is that information about the
server's SYN backlog state is entangled with information about the client's
IPID field.  Thus, we can perform both types of idle scans (IPID and SYN
backlog), to detect drops in both directions, and our technique overcomes the
limitations of both by exploiting the entanglement of information in the IPID
and treating it as a linear intervention problem to handle noise characteristic
of the real Internet. 

This research has several major contributions:

\begin{itemize}
\item A non-intrusive method for detecting intentional packet drops between two
IP addresses on the Internet where neither is a measurement machine.
\item An Internet measurement study that shows the efficacy of the method.
\item A model of IPID noise based on an autoregressive-moving-average
(ARMA) model that is robust to autocorrelated noise. 
\end{itemize}

Source code and data are available upon request, and a web demonstration
version of the hybrid idle scan is at \url{http://spookyscan.cs.unm.edu}.  The
types of measurements we describe in this paper raise ethical concerns because
the measurements can cause the appearance of connection attempts between
arbitrary clients and servers.  In China there is no evidence of the owners of
Internet hosts being persecuted for attempts to connect to the Tor network,
thus our measurements in this paper are safe.  However, we caution against
performing similar measurements in other countries or contexts without first
evaluating the risks and ethical issues.  More discussion of ethical issues is
in Section~\ref{sec:ethics}.

The rest of the paper is structured as follows: After describing the
implementation of our method in Section~\ref{sec:implementation}, we present
our experimental methodology for the measurement study in
Section~\ref{sec:experimental}.  This is followed by
Section~\ref{sec:analysis}, which describes how we analyze the time series data
generated by a scan using an ARMA model.  Results from the measurement study
are in Section~\ref{sec:results}, followed by discussions of related work in
Section~\ref{sec:relatedwork} and ethical issues in Section~\ref{sec:ethics},
and then the conclusion.

\section{Implementation} \label{sec:implementation}
\begin{figure*}[t!]
  \centering
  \includegraphics[width=0.99\textwidth]{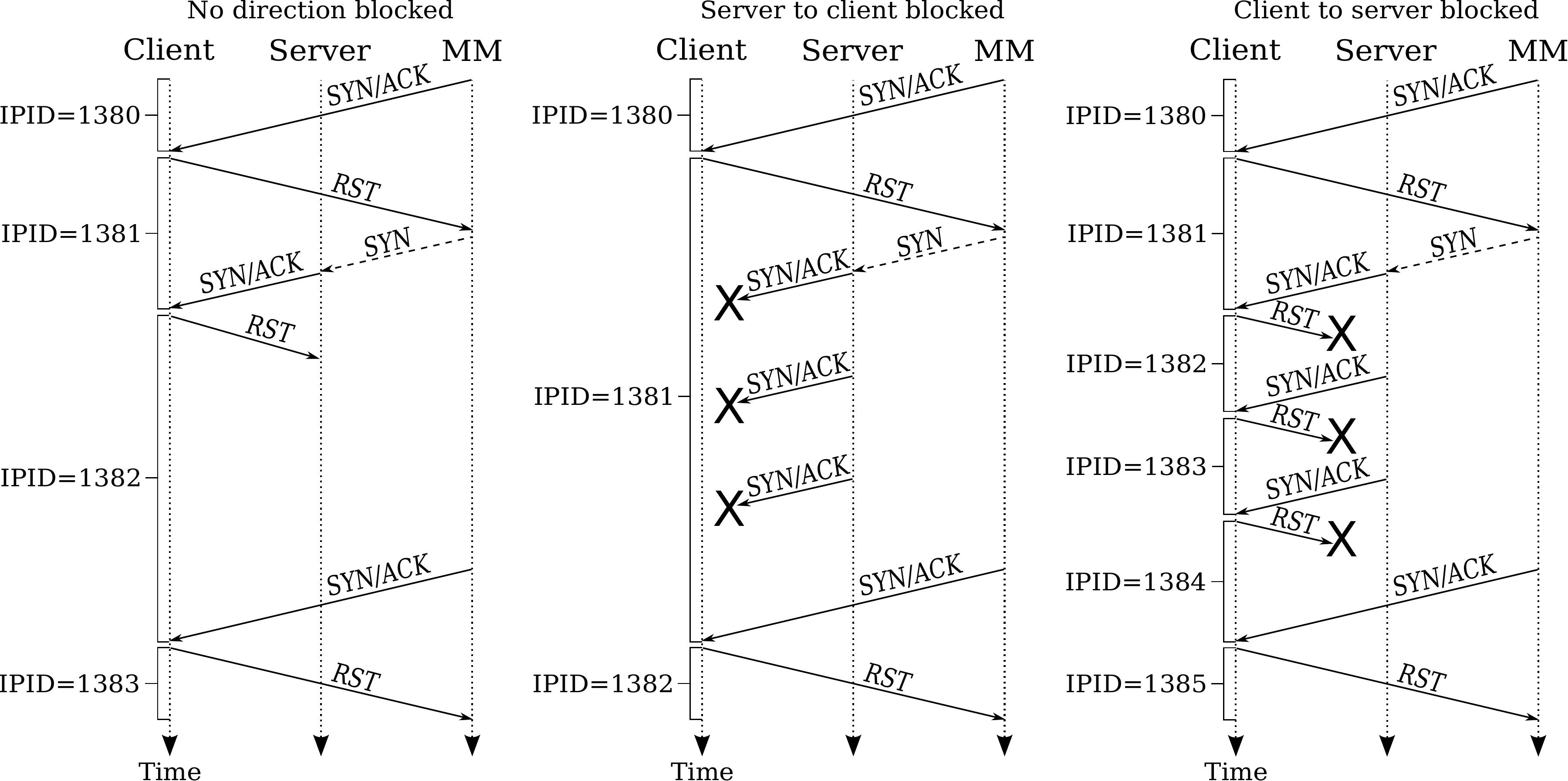}
  \caption{Three different cases that our method can detect.  MM is the measurement machine.\label{fig:threecases}}
\end{figure*}

In order to determine the direction in which packets are being blocked, our
method is based on information flow through both the IPID of the client and the
SYN backlog state of the server, as shown in Figure~\ref{fig:threecases}.  
Our implementation queries the IPID of the client (by sending SYN/ACKs from the
measurement machine and receiving RST responses) to create a time series to
compare a base case to a period of time when the server is sending SYN/ACKs to
the client (because of our forged SYNs). We
assume that the client has global IPIDs and the server has an open port.

\begin{figure*}[t!]
  \centering
  \includegraphics[width=0.8\textwidth]{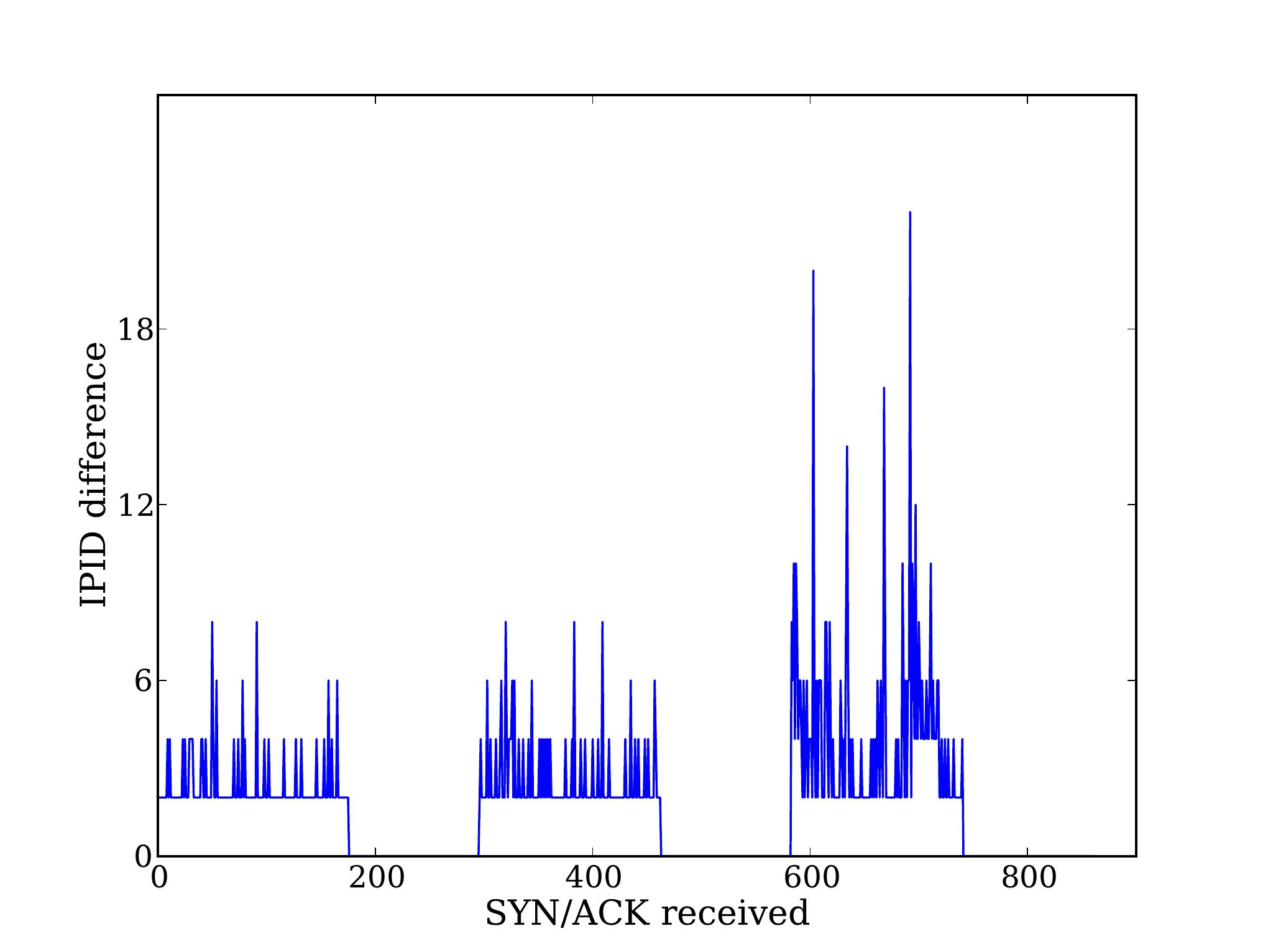}
  \caption{Example IPID difference time series' for three separate experiments
that lead to detection of the {\bf Server-to-client-dropped} case.  Note the
high amount of noise in the third experiment.  Our ARMA modeling is able to
detect this case correctly even in the presence of such high
noise.\label{fig:case1}}
\end{figure*}

\begin{figure*}[t!]
  \centering
  \includegraphics[width=0.8\textwidth]{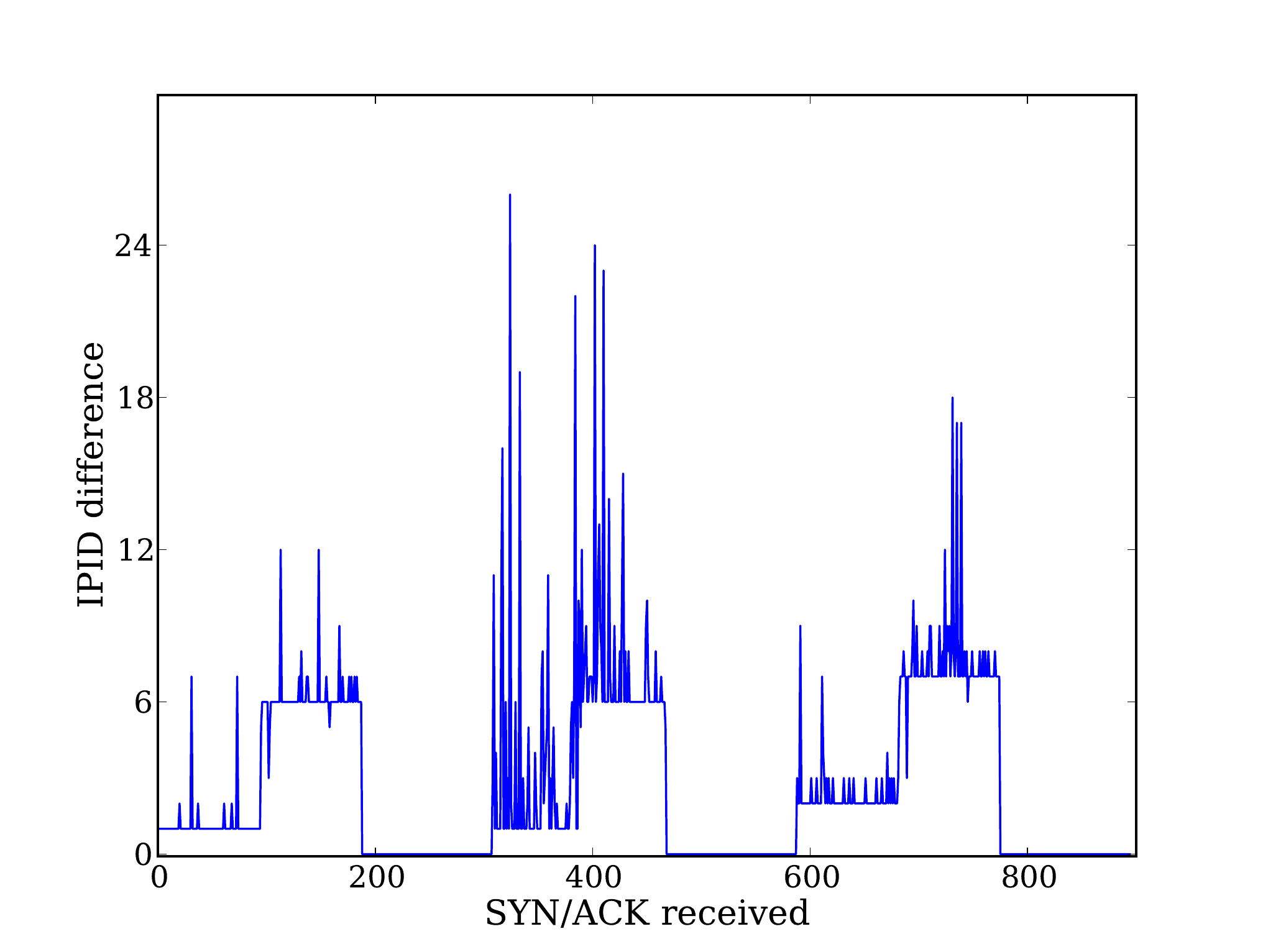}
  \caption{Example IPID difference time series' for three separate experiments
that lead to detection of the {\bf No-packets-dropped} case.  Note the high
amount of noise in the second experiment.  Our ARMA modeling is able to detect
this case correctly even in the presence of such high noise.\label{fig:case2}}
\end{figure*}

\begin{figure*}[t!]
  \centering
  \includegraphics[width=0.8\textwidth]{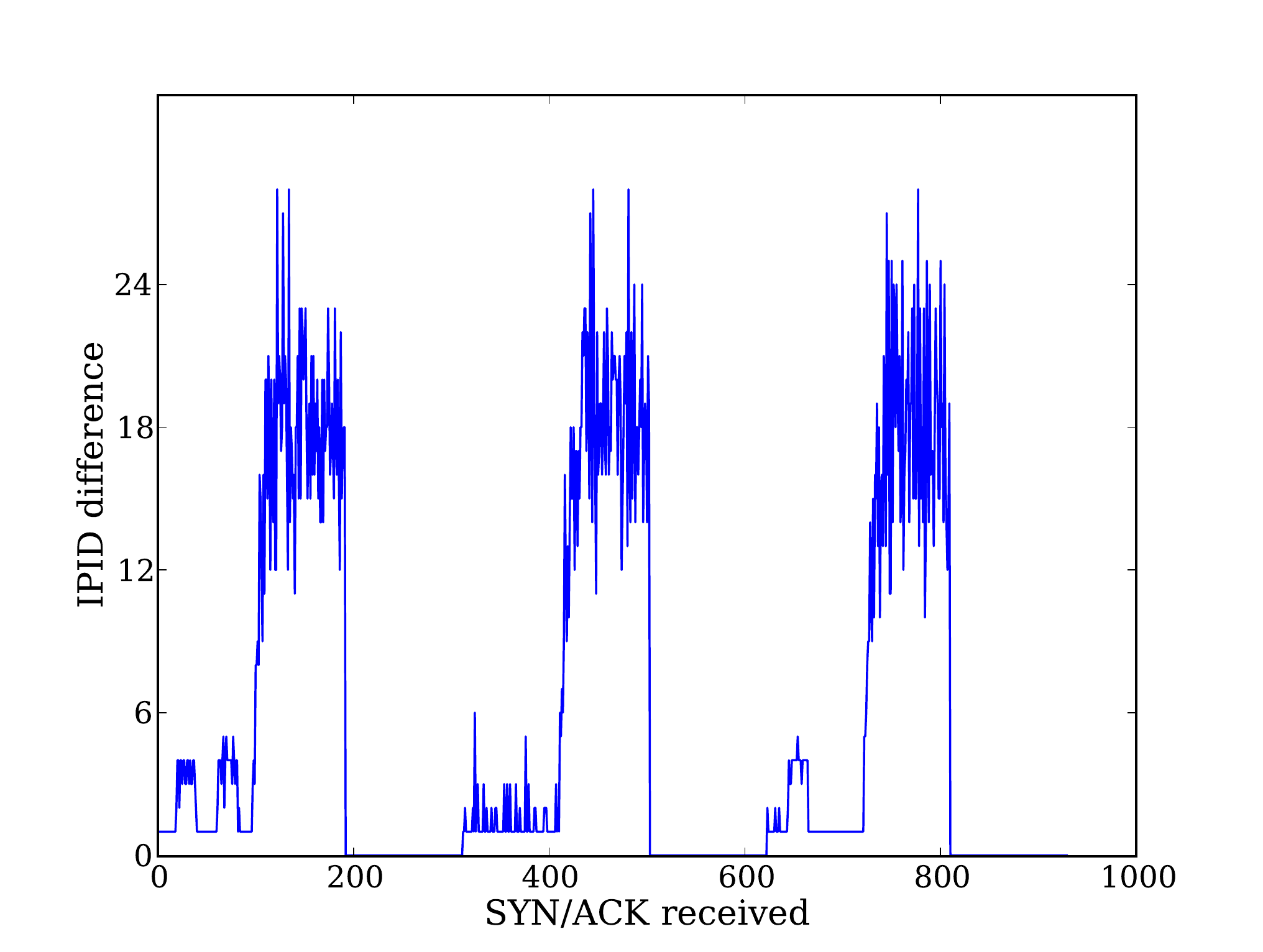}
  \caption{Example IPID difference time series' for three separate experiments
that lead to detection of the {\bf Client-to-server-dropped}
case.\label{fig:case3}}
\end{figure*}

Global IPIDs were explained in Section~\ref{sec:intro}.  The SYN backlog is a
buffer that stores information about half-open connections where a SYN has been
received and a SYN/ACK sent but no ACK reply to the SYN/ACK has been received.
Half-open connections remain in the SYN backlog until the connection is
completed with an ACK, aborted by a RST or ICMP error, or the half-open
connection times out (typically between 30 and 180 seconds).  The SYN/ACK is
retransmitted some fixed number of times that varies by operating system and
version, typically three to six SYN/ACKs in total.
This SYN backlog behavior on the server, when combined with the global IPID
behavior of the client, enables us to distinguish three different cases (plus
an error case):

\begin{packed_item}
\item {\bf Server-to-client-dropped:} In this case SYN/ACKs are dropped in
transit from the server to the client based on the return IP address (and possibly other fields like source port), and the
client's IPID will not increase at all (except for noise).  See Figure~\ref{fig:case1}.
\item {\bf No-packets-dropped:} In the case that no intentional dropping of
packets is occurring, the client's IPID will go up by exactly one.  See
Figure~\ref{fig:case2}.  This happens because the first SYN/ACK from the server
is responded to with a RST from the client, causing the server to remove the
entry from its SYN backlog and not retransmit the SYN/ACK.  Censorship that is
stateful or not based solely on IP addresses and TCP port numbers may be
detected as this case, including filtering aimed at SYN packets only.  Also, if
the packet is not dropped, but instead the censorship is based on injecting
RSTs or ICMP errors, it will be detected as this case.  Techniques for
distinguishing these other possibilities are left for future work.
\item{\bf Client-to-server-dropped:} In this case RST responses from the client
to the server are dropped in transit because of their destination IP address
(which is the server).  When this happens the server will continue to
retransmit SYN/ACKs and the client's IPID will go up by the total number of
transmitted SYN/ACKs including retransmissions (typically three to six).  See
Figure~\ref{fig:case3}.  This may indicate the simplest method for blacklisting
an IP address: null routing. 
\item{\bf Error:} In this case networking errors occur during the experiment,
the IPID is found to not be global throughout the experiment, a model is fit to
the data but does not match any of the three non-error cases above, the data is
too noisy and intervention analysis (see Section~\ref{sec:analysis}) fails
because we are not able to fit a model to the data, and/or other errors. 
\end{packed_item}

Noise due to packet loss and delay or the client's communications with other
machines may be autocorrelated.  The autocorrelation comes from the fact that
the sources of noise, which include traffic from a client that is not idle,
packet loss, packet reordering, and packet delay, are not memoryless processes
and often happen in spurts.  The accepted method for performing linear
intervention analysis on time series data with autocorrelated noise is ARMA
modeling~\cite{armabook}, which we describe in Section~\ref{sec:analysis}.

\section{Experimental Setup} \label{sec:experimental}

All measurement machines were Linux machines connected to a research network
with no packet filtering.  Specifically, this network has no stateful firewall
or egress filtering for return IP addresses.


One measurement machine was dedicated to developing a pool of both client and
server IP addresses that have the right properties for use in measurements.
Clients were chosen by horizontally scanning China and other countries for
machines with global IPIDs, then continually checking them for a 24-hour period
to cull out IP addresses that frequently changed global IPID behavior
(\emph{e.g.}, because of DHCP), went down, or were too noisy.  A machine is
considered to have a global IPID if its IPID as we measure it by sending
SYN/ACKs from alternating source IP addresses and receiving RSTs never
incrementing outside the ranges $[-40, 0)$ or $(0, 1000]$ per second when probed
from two different IP addresses.  This range allows for non-idle clients,
packet loss, and packet reordering.  It is possible to build the time series in
different ways where negative IPID differences are never observed, but in this
study our time series was the differences in the client's IPIDs in the order in
which they arrived at the measurement machine.  Our range of $[-40, 0)$ or $(0,
1000]$ is based on our observations of noise typical of the real Internet.  The
IPID going up by $0$ is a degenerate case and means the IPID is not global.

Servers were chosen from three groups: Tor directory authorities, Tor bridges,
and web servers.  The ten Tor directory authorities were obtained from the Tor
source code and the same ten IPs were tested for every day of data.  Three Tor
bridges were collected daily both through email and the web.  Every day seven
web servers were chosen randomly from the top 1000 websites on the Alexa Top
1,000,000 list~\cite{alexa}.  All web server IPs were checked to make sure that
they stood up for at least 24 hours before being selected for measurement.
Furthermore, we checked that the client and server were both up and behaving as
assumed between every experiment (\emph{i.e.}, every five minutes).

A round of experiments was a 24-hour process in which measurements were carried
out on the two measurement machines.  Each 24-hour period had 22 hours of
experiments and 2 hours of down time for data synchronization.  For each
measurement period on each of the two machines performing direct measurements,
ten server machines and ten client machines from the above process were chosen
for geographic diversity: 5 from China, 2 from countries in Asia that were not
China, 1 from Europe, and 2 from North America.  IP addresses were never reused
except for the Tor directory authorities, so that every 24-hour period was
testing a set of 20 new clients, 10 new servers, and the 10 directory
authorities. 

For each of the twenty clients and twenty servers geographical information
provided by MaxMind was saved. MaxMind claims an accuracy of 99.8\% for identifying
the country an IP address is in~\cite{maxmindaccuracy}.  For each of the twenty
server machines, a series of SYN packets was used to test and save its SYN/ACK
retransmission behavior for the analysis in Section~\ref{sec:analysis}.

Every hour, each of our two main measurement machines created ten threads.
Each thread corresponded to one client machine.  Each thread tested each of the
ten server IP addresses sequentially using our idle scan based on the client's
IPID.  No forged SYNs were sent to the server during the first 100 seconds of a
test, and forged SYNs with the return IP address of the client were sent to the
server at a rate of 5 per second for the second 100-second period.  Then forged
RST packets were sent to the server to clear the SYN backlog and prevent
interference between sequential experiments.  A timeout period of sixty seconds
was observed before the next test in the sequence was started, to allow all
other state to be cleared.  Each experiment lasted for less than five minutes,
so that ten could be completed in an hour.  Every client and server was
involved in only one experiment at a time.  Each client/server pair was tested
once per hour throughout the 24-hour period, for replication and also to
minimize the effects of diurnal patterns.  Source and destination ports for all
packets were carefully chosen and matched to minimize assumptions about what
destination ports the client responds on.  Specifically, source ports for SYN
packets sent to the server (both forged SYNs and SYNs with the measurement
machine's IP address as the return IP address for testing) were chosen from the
same range as the destination ports for SYN/ACKs send to the client (always
strictly less than 1024).  We did not find it necessary to hold the source port
for SYN/ACKs sent to the client to be always equal to the open port on the
server, but this is possible.

\section{Analysis} \label{sec:analysis}
In this section, we set out our statistical model for our time series data.
We then describe our process for outlier removal and for statistically
testing if and in which direction packet drops are occurring.

We model each time series $y_1, \ldots, y_n$ as a \emph{linear regression with
ARMA errors}, a combination of an autoregressive-moving-average (ARMA) model
with external linear regressors.  ARMA models are used to analyze time series
with autocorrelated data and are themselves a combination of two models, an
autoregressive (AR) model and a moving-average (MA) model.

An AR model of order $p$ specifies that every element of a time series can be
written as a constant plus the linear combination of the previous $p$ elements:
\[
y_t = c + z_t + \phi_1 y_{t-1} + \cdots + \phi_{t-p} y_{t-p}
\]
where $z_t$ is a white noise series.  An MA model of order $q$ specifies that
every element of a time series can be written as a constant plus the linear
combination of the previous $q$ white-noise terms:
\[
y_t = c + z_t + \theta_1 z_{t-1} + \cdots + \theta_{t-q} z_{t-q}
\]
Intuitively, each element is linearly related to the previous random ``shocks''
in the series.  An ARMA($p$, $q$) model combines an AR model of order $p$ and
an MA model of order $q$:
\[
y_t = c + z_t + \sum_{i=1}^p\phi_i y_{t-i} + \sum_{i=1}^q\theta_i z_{t-i}
\]

We use a linear regression with ARMA errors to model our time series data.
This specifies that every element in a time series can be written as a constant
plus the linear combination of regressors $x_1, \ldots, x_r$ with an
ARMA-modeled error term:
\begin{align*}
y_t &= c + \sum_{i=1}^r\beta_i x_{it} + e_t,\\
e_t &= z_t + \sum_{i=1}^p\phi_i e_{t-i} + \sum_{i=1}^q\theta_i z_{t-i}
\end{align*}

We use the regressors $x_i$ for \emph{intervention analysis}, \emph{i.e.}, for
analyzing our experimental effect on the time series at a specific time.

\begin{figure*}[t!]
  \centering
  \includegraphics[width=0.95\textwidth]{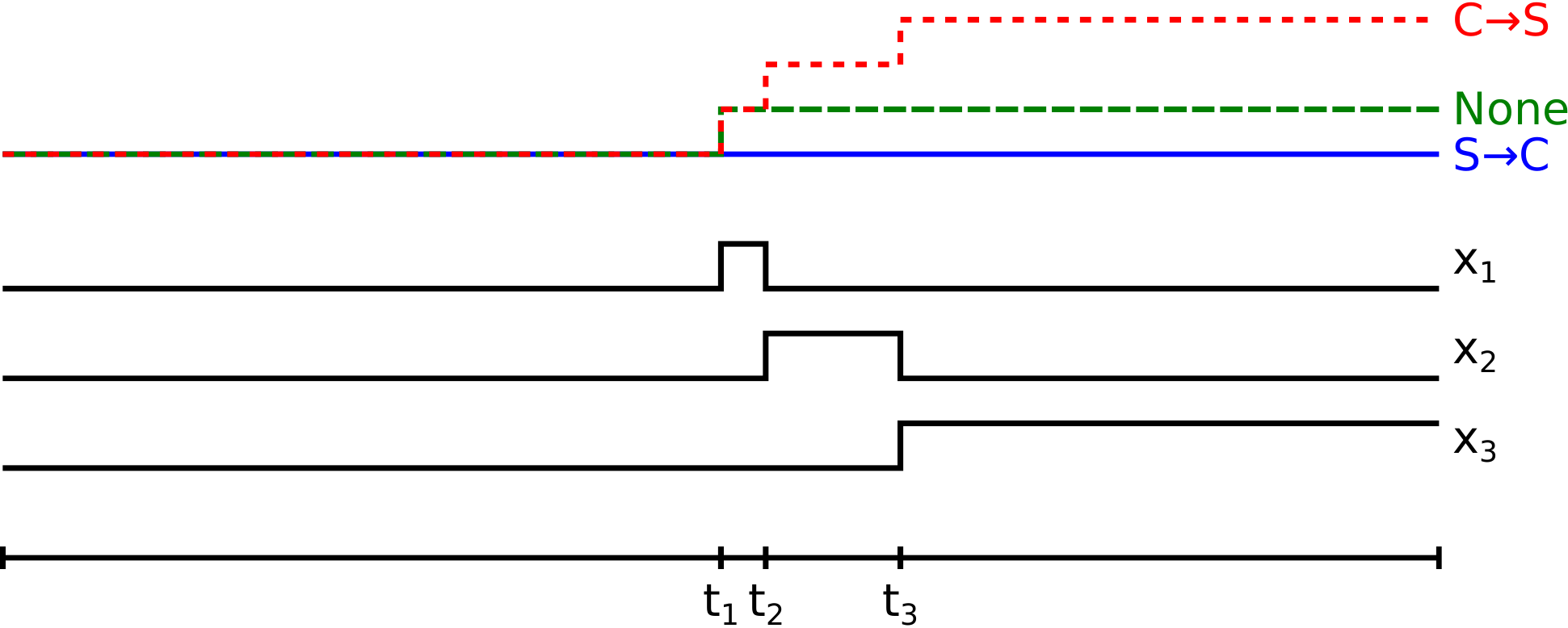}
  \caption{{\bf For a server that retransmits $r-1$ SYN/ACK's, each case can be
expressed as the linear combination of regressors $x_1, \ldots, x_r$; shown is
when $r=3$ with SYN/ACK transmissions responding to the first forged SYN
occurring at $t_1$, $t_2$, and $t_3$.  C$\rightarrow$S indicates
client-to-server, and S$\rightarrow$C indicates
server-to-client.}\label{fig:model}} \end{figure*}

For each experiment, we pick regressors according to which times the server
(re)transmits SYN/ACK's in response to SYN's.  For a server that (re)transmits
$r$ SYN/ACK's in response to each SYN, we have $r$ regressors.  We call time
$t_1$ the time of the first transmission in response to the first of our forged
SYN's, and we call $t_{i+1}$ the time the server would send the $i$th
retransmission in response to that SYN.  Then we define regressor $x_i$ as the
indicator variable
\begin{align*}
x_{ij} =
\begin{cases}
1 &\textnormal{if $t_i \le j$ and either $j < t_{i+1}$ or $i = r$}\\
0 &\textnormal{otherwise}
\end{cases}
\end{align*}
In other words, $x_1$ is zeros until the time the server transmits the first
SYN/ACK then ones until the server begins retransmitting SYN/ACK's.  The
remaining $x_i$ are zeros until the time the server would begin retransmitting
its $i$th SYN/ACK then ones until if/when the $(i+1)$th SYN/ACK's would begin
being retransmitted.  This definition allows us to model any of the possible
level shifts in any case of packet drop as a linear combination of all $x_i$.
See Figure~\ref{fig:model} for an illustration.

We choose ARMA orders $p$ and $q$ by performing model selection over time
series elements $y_1, \ldots, y_{t_1}$.  We find the $p\le7$ and $q\le7$
for the ARMA($p$, $q$) model that maximizes the corrected Akaike information
criterion, a metric which rewards models that lose less information but
penalizes models overfitted with too many parameters~\cite{aicc}.  It is given
by
\[
AIC_C = -2\ln{L} + 2k + \frac{2k(k+1)}{n-k-1},
\]
where here the number of parameters $k$ is $p + q + 2$ and where $L$ is the
estimated maximum likelihood over all $\phi_i$ and $\theta_i$.

After $p$ and $q$ are chosen, we then simultaneously fit all $\phi_i$,
$\theta_i$, and $\beta_i$ of our linear regression model with ARMA errors over
the entire time series $y_1, \ldots, y_n$ corresponding to the estimated
maximum likelihood.

After fitting parameters, we remove outliers that might be caused by,
\emph{e.g.}, spikes in network traffic that may hamper our analysis.  We use
the $\hat{\lambda}_{2,T}$ test statistic proposed by Chang \emph{et.\
al}~\cite{outliers} with significance $\alpha = 0.05$.  After removing
outliers, we iteratively refit the $\phi_i$, $\theta_i$, and $\beta_i$
parameters and test for outliers until no additional outliers are removed.

For intervention analysis, we use hypothesis testing over the value of
$\beta_r$ to determine if packets are dropped and in which direction.  If we
send $s$ forged SYN's, without noise, we would expect $\beta_r$ to equal one of
the following: $0$ for the case where packets are dropped from the server to
client, $s$ for the case where no packets are dropped, or $rs$ for the case
where packets are dropped from the client to server.  One might pick two
thresholds, $k_1 = s/2$ in between the first two cases and threshold $k_2 = (1
+ r)s/2$ between the last two cases; however, for the second threshold, we
instead choose $k_2' = \min(2s, k_2)$ to be more robust to, \emph{e.g.}, packet
loss.  Then we determine the case by a series of one-sided hypothesis tests
performed with significance $\alpha = 0.01$ according to the following
breakdown:

\begin{packed_item}
\item
\textbf{Server-to-client-dropped} if we reject the null hypothesis that
$\beta_r \ge k_1$.
\item
\textbf{No-packets-dropped} if we reject the null hypotheses that $\beta_r \le
k_1$ and that $\beta_r \ge k_2'$.
\item
\textbf{Client-to-server-dropped} if we reject the null hypothesis that
$\beta_r \le k_2'$.
\item \textbf{Error} if none of the above cases can be determined.
\end{packed_item}

\section{Results} \label{sec:results}

Table~\ref{tab:results} shows results from 5 days of data collection, where $S
\rightarrow C$ is {\bf Server-to-client-dropped}, {\it None} is {\bf
No-packets-dropped}, $C \rightarrow S$ is {\bf Client-to-server-dropped}, and
{\it Error} is {\bf Error}.  {\it CN} is China, {\it Asia-CN} is other Asian
countries, {\it EU} is Europe, and {\it NA} is North America.  For server
types, {\it Tor-dir} is a Tor directory authority, {\it Tor-bri} is a Tor
bridge, and {\it Web} is a web server.

\begin{table}
\centering
\begin{footnotesize}
\begin{tabular}{ | c | c | c | c | c |}
\hline
\hline
\multicolumn{1}{c}{Client,Server} & 
\multicolumn{1}{c}{$S \rightarrow C$ (\%)} & 
\multicolumn{1}{c}{None (\%)} & 
\multicolumn{1}{c}{$C \rightarrow S$ (\%)} & 
\multicolumn{1}{c}{Error (\%)} \\
\hline
\hline
CN,Tor-dir & 2200 (73.04)  & 19 (0.63) & 504 (16.73) & 289 (9.59) \\
Asia-CN,Tor-dir & 0 (0.00)  & 1171 (96.38) & 1 (0.08) & 43 (3.54) \\
NA,Tor-dir & 1 (0.07)  & 1217 (90.69) & 49 (3.65) & 75 (5.59) \\
EU,Tor-dir & 2 (0.28)  & 695 (97.89) & 2 (0.28) & 11 (1.55) \\
CN,Tor-bri & 1012 (58.91)  & 565 (32.89) & 31 (1.80) & 110 (6.40) \\
Asia-CN,Tor-bri & 0 (0.00)  & 626 (80.88) & 9 (1.16) & 139 (17.96) \\
NA,Tor-bri & 0 (0.00)  & 657 (78.21) & 30 (3.57) & 153 (18.21) \\
EU,Tor-bri & 0 (0.00)  & 313 (78.25) & 9 (2.25) & 78 (19.50) \\
CN,Web & 28 (2.15)  & 995 (76.30) & 36 (2.76) & 245 (18.79) \\
Asia-CN,Web & 1 (0.17)  & 569 (97.43) & 1 (0.17) & 13 (2.23) \\
NA,Web & 0 (0.00)  & 606 (93.37) & 0 (0.00) & 43 (6.63) \\
EU,Web & 0 (0.00)  & 305 (90.24) & 0 (0.00) & 33 (9.76) \\
\hline
All Web & 29 (1.01)  & 2475 (86.09) & 37 (1.29)  & 334 (11.62) \\
All Tor-bri & 1012 (27.12)  & 2161 (57.90) & 79 (2.12)  & 480 (12.86) \\
All Tor-dir & 2203 (35.09)  & 3102 (49.40) & 556 (8.85)  & 418 (6.66) \\
\hline
\end{tabular}
\caption{Results from the measurement study.\label{tab:results}}
\end{footnotesize}
\end{table}

Our expectation would be to observe {\bf Server-to-client-dropped} for
clients in China and Tor servers because of Winter and Lindskog's observation
that the SYN/ACKs are statelessly dropped by the ``Great Firewall of China''
(GFW) based on source IP address and port~\cite{winter2012}.  We would expect
to see {\bf No-packets-dropped} for most web servers from clients in China,
unless they host popular websites that happen to be censored in China.
Similarly, in the expected case we should observe {\bf No-packets-dropped} for
clients outside of China, regardless of server type.  We expect a few
exceptions, because censorship happens outside of China and because
the GFW is not always 100\% effective.  In particular, Tor bridges are not
blocked until the GFW operators learn about them, and some routes might not
have filtering in place.  Our results are congruent with all of these
expectations. 

In 5.9\% of the client/server pairs we tested, multiple cases were observed in
the same day.  In some cases it appears that noise caused the wrong case to be
detected, but other cases may be attributable to routes changing throughout the
day~\cite{tcp5}.  That the data is largely congruent with our expectations
demonstrates the efficacy of the approach, and some of the data points that lie
outside our expectations have patterns that suggest that a real effect is being
measured, rather than an error.  For example, of the 28 data points where web
servers were blocked from the server to the client in China, 20 of those data
points are the same client/server pair. 

38\% of the data we collected does not appear in Table~\ref{tab:results}
because it did not pass liveness tests.  Every 5-minute data point has three
associated liveness tests.  If a server sends fewer than 2.5 SYN/ACKs in
response to SYNs from the measurement machine, a client responds to less than
$\frac{3}{5}$ of our SYN/ACKs, or a measurement machine sending thread becomes
unresponsive, that 5-minute data point is discarded.

Two out of the ten Tor directory authorities never retransmitted enough
SYN/ACKs to be included in our data.  Of the remaining eight, two more account
for 98.8\% of the data points showing blocking from client to server.
These same two directory authorities also account for 72.7\% of the {\bf Error}
cases for directory authorities tested from clients in China, and the case of
packets being dropped from server to client (the expected case for China and
the case of the majority of our results) was never observed for these two
directory authorities.

When Winter and Lindskog~\cite{winter2012} measured Tor reachability from a
virtual private server in China, there were eight directory authorities at that
time.  One of the eight was completely accessible, and the other seven were
completely blocked in the IP layer by destination IP (\emph{i.e.}, {\bf
Client-to-server}).  In our results, six out of ten are at least blocked {\bf
Server-to-client} and two out of ten are only blocked {\bf Client-to-server}
(two had all results discarded).  Winter and Lindskog also observed that Tor
relays were accessible 1.6\% of the time, and we observed that directory
authorities were accessible 0.63\% of the time.  Our results have geographic
diversity and their results can serve as a ground truth because they tested
from within China.  In both studies the same special treatment of directory
authorities compared to relays or bridges was observed, as well as a small
percentage of cases where filtering that should have occurred did not.

To evaluate the assumption that clients with a global IPID are easy to find in
a range of IP addresses that we desire to measure from, take China as an
example.  On average, 10\% of the IP addresses in China responded to our probes
so that we could observe their IPID, and of those 13\% were global.  So,
roughly 1\% of the IP address space of China can be used as clients for
measurements with our method, enabling experiments with excellent geographic
and topological diversity.

\section{Related Work} \label{sec:relatedwork}

Related work directly related to idle scans~\cite{antirez,nmapbook,roya} was
discussed in Section~\ref{sec:intro}.  Other advanced methods for inferring
remote information about networks have been proposed.  Qian \emph{et
al.}~\cite{Qian:2012:OTS:2310656.2310690}
demonstrate that firewall behavior with respect to sequence numbers can be used
to infer sequence numbers and perform off-path TCP/IP connection hijacking.
Chen \emph{et al.}~\cite{Chen:2005:EIF:2150193.2150205} use the IPID field to
perform advanced inferences about the amount of internal traffic generated by a
server, the number of servers in a load-balanced setting, and one-way delays.
Morbitzer~\cite{morbitzerthesis} explores idle scans in IPv6.

iPlane~\cite{Madhyastha:2006:IIP:1298455.1298490} sends packets from PlanetLab
nodes to carefully chosen hosts, and then compounds loss on specific routes to
estimate the packet loss between arbitrary endpoints without access to those
endpoints.  This does not detect IP-address-specific packet drops.  Our
technique, in contrast, can be used to detect intentional drops of packets
based on IP address and requires no commonalities between the measurement
machine's routes to the server or client and the routes between the server and
client.  Queen~\cite{Wang:2009:QEP:1532940.1532949} utilizes recursive DNS
queries to measure the packet loss between a pair of DNS servers, and
extrapolates from this to estimate the packet loss rate between arbitrary
hosts.  Reverse traceroutes can be performed by forging return IP addresses and
using the IP options for recording routes and
timestamps~\cite{Katz-Bassett:2010:RT:1855711.1855726}.
De A. Rocha \emph{et al.}~\cite{DeA.Rocha:2007:NAM:1772322.1772439}
present a method for estimating average variance and delay based on forged
return IP addresses and the IPID field.

Dainotti \emph{et al.}~\cite{Dainotti:2011:ACI:2068816.2068818} analyze several
Internet disruption events that were censorship-related using various data
sources from both the control and data planes.  Flach \emph{et
al.}~\cite{Flach:2012:QVD:2398776.2398804} present a method for detecting cases
where routing decisions are not solely based on destination IP address.  In
general, understanding reachability and routing issues on the Internet is an
important problem, and we assert that idle scans are a promising general
approach to perform these kinds of measurements.

\section{Discussion of Ethical Issues} \label{sec:ethics}

The main ethical concerns with the measurements presented in this paper arise
from the fact that we are essentially creating traffic between a client and
server, where the client is typically inside a domain where access to the
server might be blocked.  Simply creating such traffic may have negative
consequences for the owner/operator of the client machine.  A separate ethical
concern is raised by the measurements themselves, because we are sending SYN
packets at a relatively high rate, some of them forged, with no intention of
completing a connection with the server.

Regarding the creation of traffic between the client and server, assuming that
the path from the measurement machine to the server does not go through the
censor's Internet infrastructure, then the traffic generated between the client
and server that the censor can see is only SYN/ACKs from the server to the
client and RSTs from the client to the server.  Nonetheless, if this
information is reported to authorities in an aggregated form, such as netflow
records or aggregate bandwidth numbers, then it could appear that the client is
communicating with the server.  Thus we strongly recommend that the
measurements we present in this paper not be carried out without a full
understanding of the context and ethical considerations specific to the country
being studied.  China has no history of persecuting Internet users for
attempting to connect to evasion technologies such as the Tor network.  China's
basic approach to censorship and surveillance on the Internet is to have these
functions carried out by companies (see, \emph{e.g.}, Crandall \emph{et
al.}~\cite{FM4628}), with the government only stepping in when the companies
fail to do an adequate job~\cite{anaconda}.  For more information about
Internet controls in China, see the Open Net Initiative's country
profile~\cite{onichina}.

Regarding the fact that we are sending SYN packets to the server at a
relatively high rate with no intention of completing a connection, note that
the rate we are sending SYN packets (5 per second) is not enough to create a
denial-of-service condition on any modern network stack.  Modern network stacks
have mechanisms for preventing their SYN backlogs from being filled except at
very high rates, and, even if the SYN backlog is full, modern operating systems
typically have SYN cookies~\cite{syncookies} turned on by default.  Causing the
SYN backlog to fill is still a potential denial-of-service when SYN cookies are
enabled, because SYN cookies are never retransmitted and often exclude
important features such as the scaled flow control window.  Fortunately, our
hybrid idle scan, unlike the SYN backlog idle scan of Ensafi \emph{et
al.}~\cite{roya}, does not require the SYN backlog to be full before
information is leaked about the SYN backlog state.  For an interesting
discussion about ethical issues related to port scans in general, we refer the
reader to Durumeric \emph{et al.}~\cite{zmap13}.

\section{Conclusion} \label{sec:conclusion}

We have presented a method for detecting intentional packet drops (\emph{e.g.},
due to censorship) between two almost arbitrary hosts on the Internet, assuming
the client has a globally incrementing IPID and the server has an open port.
Our method can determine which direction packets are being dropped in, and is
resistant to noise due to our use of an ARMA model for intervention analysis.
Our measurement results are congruent with current understandings about global
Internet censorship, demonstrating the efficacy of the method.

\section{Acknowledgments}

We would like to thank the anonymous PAM 2014 reviewers and our shepherd,
Jelena Mirkovic, as well as Terran Lane, Patrick Bridges, Michalis Faloutsos,
Stefan Savage, and Vern Paxson for helpful feedback on this work.  This
material is based upon work supported by the National Science Foundation under
Grant Nos. \#0844880, \#1017602, \#0905177, and \#1314297.
\bibliographystyle{splncs}
\bibliography{references}

\end{sloppypar}
\end{document}